\documentstyle[12pt]{article}
\input epsf

\begin{document}
\baselineskip=15pt
\newcommand{\x}{{\bf x}}
\newcommand{\y}{{\bf y}}
\newcommand{\z}{{\bf z}}
\newcommand{\bp}{{\bf p}}
\newcommand{\A}{{\bf A}}
\newcommand{\B}{{\bf B}}
\newcommand{\p}{\varphi}
\newcommand{\hp}{\hat\varphi}
\newcommand{\del}{\nabla}
\newcommand{\be}{\begin{equation}}
\newcommand{\ee}{\end{equation}}
\newcommand{\bq}{\begin{eqnarray}}
\newcommand{\eq}{\end{eqnarray}}
\newcommand{\ba}{\begin{eqnarray}}
\newcommand{\ea}{\end{eqnarray}}
\def\r{\nonumber\cr}
\def\hf{\textstyle{1\over2}}
\def\qr{\textstyle{1\over4}}
\def\Sc{Schr\"odinger\,}
\def\sc{Schr\"odinger\,}
\def\'{^\prime}
\def\>{\rangle}
\def\<{\langle}
\def\-{\rightarrow}
\def\dbd{\partial\over\partial}
\def\tr{{\rm tr}}
\def\hg{{\hat g}}
\def\ca{{\cal A}}
\def\pd{\partial}
\def\dl{\delta}

\begin{titlepage}
\vskip1in
\begin{center}
{\large AdS/CFT boundary conditions, multi-trace perturbations,
and the c-theorem. }
\end{center}
\vskip1in
\begin{center}
{\large David Nolland}

\vskip20pt

Department of Mathematical Sciences

University of Liverpool

Liverpool, L69 3BX, England

{\it nolland@liv.ac.uk}

\end{center}
\vskip1in
\begin{abstract}

\noindent We discuss possible choices for boundary conditions in
the AdS/CFT correspondence, and calculate the renormalisation
group flow induced by a double-trace perturbation. In running from
the UV to the IR there is a unit shift in the central charge. The
discrepancy between our result and results obtained by other
authors is accounted for by the discovery that there is a
non-trivial flow for perturbations induced by bulk fields with
masses saturating the Breitenlohner-Freedman bound.

\end{abstract}

\end{titlepage}

\section{Introduction}

The AdS/CFT correspondence \cite{review}, in relating conventional
quantum gauge field theories to gravitational and string theories
in higher dimensions, has proved to be of great importance in
elucidating non-perturbative aspects of both kinds of theory, and
is likely to retain a central role for some time to come. Most of
the work that has been done on this correspondence has involved
taking the large-$N$ limit of the gauge theory, which corresponds
to the classical limit of the gravity theory. But it is of
considerable interest to go beyond this limit and consider loop
corrections on the gravity side, which give $O(1/N)$ corrections
to the gauge theory.

The calculation of string loops in AdS backgrounds is difficult,
because the cancellation of divergences is not well understood,
and the Ramond-Ramond fields make calculations in the string genus
expansion largely intractable. However, for the calculation of
certain quantities it is sufficient to consider loop corrections
in the Supergravity limit of string theory. This is particularly
true when there are non-renormalisation theorems protecting the
quantities on the gauge theory side.

It is also interesting to consider relevant perturbations of the
gauge theory that correspond to tachyonic fields in the
Supergravity theory. These break the conformal symmetry of the
boundary gauge theory, and drive a renormalisation group flow. To
understand the effects of these perturbations on the gauge theory
requires an understanding of their asymptotic behaviour near the
boundary, and the boundary conditions that we can impose on them.
These boundary conditions were first considered in
\cite{witten,berkooz}, though the older work of \cite{bf} is also
relevant. Boundary condtions were also discussed in \cite{minces}.
For tachyonic modes whose masses lie in an appropriate range, the
difference between the ultraviolet and infrared fixed points is
just a difference in boundary conditions. We will describe these
boundary conditions from a Hamiltonian perspective in which this
difference corresponds to the choice of Dirichlet or Neumann
boundary conditions for the bulk field. The partition functions of
the perturbed boundary theory are then related by a functional
Fourier transform (in the large-N limit a saddle point
approximation reduces this to a Legendre transform).

In a series of publications we have obtained results on one-loop
Weyl anomalies in AdS/CFT \cite{us,us2,us3} that reproduce the
exact form of the anomaly on the gauge theory side, including 1/N
corrections. The coefficients of these anomalies are central
charges that, according to the holographic c-theorem, should be
larger for the infrared fixed point than the ultraviolet one, for
the range of scaling dimensions where both correspond to
normalisable perturbations. The main purpose of this paper is to
verify this using our calculation of the Weyl anomaly.

The result we obtain is in contradiction with the work of
\cite{gubser}, which was made use of in other studies of the
renormalisation group flow in AdS and dS spaces \cite{odintsov}.
The calculation of \cite{gubser} assumed that there is no flow for
bulk masses saturating the Breitenlohner-Freedman bound. We will
show that this is not true; there is a non-trivial flow in this
case that can be related to an ambiguity in the boundary term
\cite{dobrev,kw}. Hopefully this accounts for the discrepancy with
our result for the anomaly, which is linear in the scaling
dimension of the field.

This paper is organised as follows: in Section 2 we discuss
boundary conditions for AdS fields. In Section 3 we discuss
canonical quantisation. In Section 4 we discuss the AdS/CFT
correspondence formulae for single and multi-trace perturbations,
and show how they are related to boundary conditions for bulk
fields. In Section 5 we review our calculation of the Weyl anomaly
and calculate the running of the central charge for a double-trace
perturbation.

\section{Boundary Conditions on AdS Fields}

We write the metric of Euclidean $AdS_{d+1}$ as

\be ds^2={1\over z^2}\left(dz^2+\sum^{d}_{i=1}
dx_i^2\right)=dr^2+z^{-2}\sum^{d}_{i=1} dx_i^2, \ee

where $z=\exp(r)$, and we have set the length scale of AdS to
unity. The boundary is at $z=0$. Now consider a scalar field of
mass $m$ propagating in this metric. Near the boundary it has the
asymptotic behaviour

\be \phi=\alpha(x)z^{d-\Delta}+\beta(x)z^{\Delta}+\ldots,
\label{asympt}\ee

where $\Delta$ is a root of the equation

\be \Delta(\Delta-d)=m^2\label{se}.\ee

For $\Delta=d/2$ the second asymptotic solution goes like
$z^{d/2}\ln z$. From a Hamiltonian point of view $\alpha$ and
$\beta$ are conjugate variables, in a sense that we will make more
precise shortly.

Let us consider more carefully how the asymptotic form
(\ref{asympt}) of bulk fields near the boundary is related to the
boundary condition. The action for a free scalar field in AdS is
given by

\be \int d^{d+1}x\sqrt g{1\over
2}\left(g^{\mu\nu}\partial_\mu\phi\partial_\nu\phi+m^2\phi^2\right),
\ee

with linear variation

\be \int d^{d+1}x\sqrt
g\delta\phi\left(-\del^2+m^2\right)\phi+{1\over2}\int d^dx
z^{-d+1}\left(\phi\partial_z\delta\phi-(\partial_z\phi)\delta\phi\right).
\ee

For this variation to vanish, we need $\phi$ to obey the classical
equations of motion in the bulk, with the boundary condition
\cite{berkooz}

\be \partial_r\phi|_{\partial AdS}=\omega\phi|_{\partial AdS}, \ee

where $\omega$ is arbitrary. If we take $\omega=\Delta$ and insert
the asymptotic form (\ref{asympt}) for $\phi$ near the boundary,
this becomes

\be z^{\Delta-d}(\partial_r-\Delta)\phi|_{\partial AdS}=0,
\label{bcs}\ee

where we multiplied by $z^{d-\Delta}$ to obtain a relation that is
finite at $z=0$. This condition diagonalises the value of
$\alpha$, but places no restriction on $\beta$. So we can identify
$\alpha$, $\beta$ as conjugate variables.

Suppose that $\Delta$ is the smaller root of (\ref{se}). If we
make the change of variables $\tilde\phi=z^{-\Delta}\phi$ (so that
$\tilde\phi\sim\beta$ at the boundary) then we see that
(\ref{bcs}) corresponds to Neumann boundary conditions for
$\tilde\phi$. If $\Delta$ is the larger root of (\ref{se}) then we
can change variables to $\bar\phi=z^{\Delta-d}\phi$ so that
$\bar\phi\sim\alpha$ at the boundary, and then (\ref{bcs})
corresponds to Dirichlet boundary conditions for $\bar\phi$.
Henceforth we will use $\Delta$ to denote the larger root of
(\ref{se}).

In the case where $\Delta=d/2$, we can change variables to
$z^{-d/2}\phi$ and impose either Dirichlet or Neumann boundary
conditions on this field, corresponding again to diagonalising the
two possible asymptotics of the field.

At the quantum level, the above boundary conditions can be imposed
by adding a boundary term to the action; these boundary terms may
be renormalised by interactions of the bulk field.

\section{Canonical Quantisation}

Consider a free scalar field of mass $m$. For greater generality
it is convenient to perturb the $AdS_{d+1}$ metric in such a way
that the boundary has a d-dimensional Einstein metric $\hat g$
\cite{us2}. The perturbed metric is

\be ds^2 = G_{\mu\nu}\,dX^\mu\,dX^\nu=dr^2 + z^{-2}¥\, e^{\rho}
\hg_{ij}(x)\, dx^i dx^j \, ,\quad e^{\rho/2}= 1-C\,z^{2}\,,
\quad¥C={l^2 {\hat R} \over 4\,d(d-1)}\,,\label{ads1} \ee

where $\hat R$ is the Ricci tensor on the boundary, and the $d+1$
dimensional Einstein equations are still satisfied. The action for
the scalar field in this metric can be written as

\bq S_\phi&=&{1\over 2}\int d^{d+1} X{\sqrt
G}\left(G^{\mu\nu}\partial_\mu\phi\,
\partial_\nu\phi+m^2\phi^2\right)\nonumber \\
&=&{1\over 2}\int {d^dx\,dr\over
z^4}{\sqrt\hg}\,e^{2\rho}\,\left(\dot\phi^2 +z^2
e^{-\rho}\hg^{ij}\partial_i\phi\,\partial_j\phi+m^2\phi^2\right)
\,,\label{sca} \eq

where the dot denotes differentiation with respect to $r$. The
norm on fluctuations of the field, from which the functional
integral volume element ${\cal D}\phi$ can be constructed is \be
||\delta\phi||^2=\int d^{d+1} X{\sqrt G}\,\delta\phi^2 =\int
{d^dx\,dr\over z^d}{\sqrt\hg}\,e^{{d\over2}\rho}\,\delta\phi^2\,.
\ee We will interpret the co-ordinate $r$ as Euclidean time, so to
write down a Schr\"odinger equation we first re-define the field
by setting $\phi=z^{d\over2}\,e^{-{d\over4}\rho}\varphi$ to make
the `kinetic' term in the action into the standard form, and
remove the explicit $r$-dependence from the integrand of the norm.
The action becomes \bq S_\phi&=&{1\over 2}\int
{d^dx\,dr}{\sqrt\hg}\,\left(\dot\varphi^2+z^2 e^{-\rho}\varphi
\left(\Box+{(d-2){\hat R}\over4(d-1)
}\right)\varphi+\left(m^2+{d^2\over
4}\right)\varphi^2\right)\nonumber\\&&-{1\over 2}\int
d^dx\,{\sqrt\hg}
\left({d\over4}\dot\rho+{d\over 2}\right)\varphi^2\,,\nonumber\\
&=&S_\varphi+S_b\,, \label{apb}\eq where $\Box$ is the
d-dimensional covariant Laplacian constructed from $\hat g$. Note
that $\Box+{(d-2){\hat R}\over4(d-1) }$ is the operator associated
with a conformally coupled d-dimensional field, and the mass has
been modified to an effective mass $M_r=\sqrt{m^2+d^2/4}$.

In the first instance we will consider diagonalising the boundary
value of $\varphi$, so that $\dot\varphi$ is represented by
functional differentiation acting on a boundary wave-functional.
This wave-functional can be represented by a path integral

\be \int {\cal
D}\phi\,e^{-S_{\phi}¥}¥\Big|_{\phi(r=r_0)=\hat\phi}¥
=e^{-S_b}\int {\cal
D}\varphi\,e^{-S_{\varphi}¥}¥\Big|_{\varphi(r=r_0)=\hat\varphi}
¥\equiv e^{-S_b+W[\hat\varphi,g]}, \qquad W[\hp]=F+\hf\int
d^dx\sqrt{\hat g}\hp\Gamma\hp,\label{wf}\ee where a regularisation
is achived by taking the boundary at $r=r_0$ rather than
$r=-\infty$.

 This satisfies a functional Schr\"odinger equation that can be read off from the action
and gives

\be {\pd \over \pd
r_0}\Gamma=\Gamma^2-\tau^{2}e^{-\rho}\left(\Box+{(d-2){\hat
R}\over4(d-1) }\right)+M_r^2,\qquad {\partial\over\partial
r_0}F=\hf{\rm Tr}\Gamma ,\label{gam} \ee

solved by expanding $\Gamma$ in powers of the differential
operator \cite{us2}. The result is easily summed in terms of
Bessel functions. To get the correct scaling dimension as
$r_0\to-\infty$ requires discarding terms of order less than
$\tau^{2M_r}$ in the asymptotic expansion of $\Gamma$. Such terms
should be removed by an appropriate renormalisation.  Namely we
discard these terms so that $\Gamma$ has the asymptotic behaviour

\be \Gamma\sim \tau^{2M_r}p^{2M_r}. \label{ag}\ee

Here $\tau=\ln r_0$ is the boundary value of $z$. To get a finite
wave-functional as the cutoff is removed, we perform a
wave-function renormalisation $\hp\rightarrow \tau^{-M_r}\hp$. The
scaling dimension of our canonical field can be read off the
asymptotic form of the wave-functional, and has the correct value
(by construction). Now $\hp$ corresponds to $\alpha$ in
(\ref{asympt}), where the scaling dimension $\Delta$ is taken to
be the larger root of (\ref{se}). This is in accordance with our
previous discovery that diagonalising $\alpha$ corresponds to a
Dirichlet condition on a suitably defined bulk field.

To get the other condition, we can perform a functional Fourier
transform on the boundary. This gives us a Neumann condition for
our "canonical" field, so that the wave-functional is written in
terms of a boundary value $\pi$ for the field $\dot\varphi$. When
the wave-function renormalisation is taken into account, it is
clear that $\pi$ must undergo a wave-function renormalisation
$\pi\rightarrow\tau^{M_r}\pi$. Then $\pi$ corresponds to the other
asymptotic in (\ref{asympt}) and has the correct scaling
dimension. Henceforth when we talk about imposing Dirichlet or
Neumann conditions on the canonical field, it is important to note
that it is the {\em wave-function renormalised} canonical field on
which Dirichlet or Neumann conditions are imposed.

In the case where $\Delta=d/2$ there is no wave-function
renormalisation, and as before we can impose either Dirichlet or
Neumann conditions on the canonical field.

The wave-functional with Dirichlet conditions is always
normalisable, but it is not immediately clear whether this is true
for the wave-functional with Neumann conditions. Unitarity
constraints on the bulk field reveal that the latter is
normalisable if and only if $M_r\le1$ \cite{kw}.

Notice that although this discussion applied to free scalar
fields, it applies equally well to interacting bulk scalars (of
course there may be additional renormalisations to deal with).
Also, it is straightforward to extend the analysis to fields of
other spin, and in this way perturbations preserving some
supersymmetry could be considered.

\section{AdS/CFT Correspondence Formulae}

In the usual version of the AdS/CFT correspondence, we equate the
partition functions of the conformal boundary theory and the bulk
gravitational theory:

\be Z[\phi]_{grav}=Z[\phi]_{CFT}\equiv\langle\exp{\int\phi{\cal
O}}\rangle_{CFT}. \label{pre}\ee

Here $\hat\phi$ on the CFT side is a source for the operator
${\cal O}$, which is any scalar primary of the boundary theory.
The partition function on the left-hand side can be identified
with the wave-functional considered in the last section. On the
gravity side it corresponds to a boundary value for the
corresponding bulk field. The usual prescription is to diagonalise
the boundary value of (\ref{asympt}) corresponding to the {\em
larger} scaling dimension, since this always gives a normalisable
solution. However, as first found in \cite{bf}, in the range
$d/2<\Delta<d/2+1$ both asymptotics in (\ref{asympt}) are
normalisable, and in this case a single bulk field gives rise to
two different relevant perturbations of the CFT.

A multi-trace interaction in the boundary theory is obtained by
adding an extra term to the boundary action:

\be I_{pert}=I_{CFT}+W[{\cal O}], \ee

where $W[{\cal O}]$ is an arbitrary functional of primary scalar
operators. Let us see how this affects the canonical quantisation.
For the purposes of illustration it is convenient to consider a
double-trace perturbation, so consider the partition function

\be Z_f[\phi]=\langle \exp(-\int{f\over2}{\cal O}^2+\int{\cal
O}\phi\rangle_{CFT}. \ee

The coupling to ${\cal O}$ can be linearised with a
Hubbard-Stratonovich transformation:

\be Z_f[\phi]=\det\,^{1/2}\left(-{1\over f}\right)\int
D\sigma\langle \exp\int\left({1\over2f}\sigma^2+(\sigma+\phi){\cal
O}\right)\rangle_{CFT}. \ee

As we have seen, according to the AdS/CFT correspondence,
$Z_0[\phi]$ can be interpreted as the wave-functional of a bulk
scalar. This implies that to quadratic order

\be Z_f[\phi]=\det\,^{1/2}\left(-{1\over f}\right)\int
D\sigma\exp\int\left(F+\hf(\phi+\sigma)\Gamma(\phi+\sigma)+{1\over2f}\sigma^2\right),\ee

where $F$ and $\Gamma$ are the free energy and quadratic kernel
given in Section 3. Performing the $\sigma$ integral, we have

\be Z_f[\phi]=\det\,^{-1/2}\left(f\Gamma+1\right)\exp\left(F+\hf
\phi{\Gamma\over1+f\Gamma}\phi\right).\label{ppf}\ee

Notice that for small $f$ we recover the previous result for
$Z_0[\phi]$, while in the limit $f\to\infty$ the kernel $\Gamma$
is replaced with $1/f$. Now consider how all this is affected by
the wave-function renormalisation. Assuming we started with
Dirichlet conditions, $\Gamma\sim\tau^{2M_r}$, $\phi$ is
renormalised by $\phi\to\tau^{-M_r}\phi$ as before, and the effect
is to send $f\to0$.

If we started with Neumann conditions on the canonical field then
$\Gamma\sim\tau^{-2M_r}$. For any $f$ except $f=0$ there is no
wave-function renormalisation: (\ref{ppf}) gives a wave-functional
that is finite as the cutoff is removed:

\be Z_f[\phi]=\det\,^{-1/2}\left(f\Gamma+1\right)\exp\left(F+\hf
\phi{1\over f}\phi\right).\label{pp}\ee

Since the conjugate field is represented on this wave-functional
by functional differentiation, we see that this corresponds to
imposing the boundary conditions $\alpha=f\beta$ on the bulk field
(\ref{asympt}), in accordance with the prescription of
\cite{witten}. Thus the limit $f\to\infty$ corresponds to Neumann
conditions on the canonical field (the kernel in (\ref{ppf})
naturally tends to zero in this limit because we did not introduce
a source for the conjugate field). As pointed out in
\cite{gubser}, only for $f=0$ or $\infty$ does the bulk propagator
respect SO(4,2) invariance. So only in these cases should we
expect the AdS/CFT correspondence to work.

In conclusion, we can identify Dirichlet and Neumann conditions on
the canonical field with the $f=0$ and $f=\infty$ limits of
(\ref{ppf}) respectively. Finite $f$ represents a mixture of
Dirichlet and Neumann conditions, but one loop effects will deform
the AdS background in this case.

From the gauge theory side the situation is as follows. The
perturbation $\hf f{\cal O}^2$ drives a renormalisation group flow
from a UV fixed point, where $f=0$, to an IR fixed point, at
$f=\infty$. At the fixed points the gravity dual of this theory
lives on AdS space, and the UV and IR fixed points correspond
resepectively to Dirichlet and Neumann conditions for the bulk
field dual to ${\cal O}$.

Multi-trace perturbations of dimension more than two are similarly
described by the prescription of \cite{witten}. If we Fourier
transform (\ref{pp}) we get a functional written in terms of a
source for the canonical conjugate field

\be
Z_f[\pi]=\det\,^{-1/2}\left((f\Gamma+1)/f\right)\exp\left(F-\hf
f\pi^2\right).\label{pp2}\ee

From this we see that as a result of the wave-function
renormalisation the prescription for a perturbation $W[{\cal O}]$
is to replace the wave-functional (\ref{pre}) with $W[\pi]$, where
$\pi$ is a boundary value for the {\em conjugate} of the canonical
field. The free energy is also changed as a result of the extra
determinant in (\ref{pp2}).

\section{Central charges and the c-theorem}

On general grounds, \cite{bonora,Duff1}, the Weyl anomaly takes
the form ${\cal A}=-a \,E-c\,I$ where $E$ is the Euler density,
$(R^{ijkl}R_{ijkl}-4R^{ij}R_{ij}+R^{2})/64$, and $I$ is the square
of the Weyl tensor,
$I=(-R^{ijkl}R_{ijkl}+2R^{ij}R_{ij}-R^{2}/3)/64$. The c-theorem in
four dimensions \cite{cardy,warner} suggests that the central
charge as defined in \cite{cardy} (which is related to heat-kernel
coefficients and is in general a combination of $a$ and $c$)
should be larger in the ultraviolet than in the infrared. In this
section we will check this for a double-trace deformation.

The exact $N$-dependence of the Weyl anomaly of the boundary CFT
was calculated from the AdS/CFT correspondence in \cite{us,us3}
(an overview of the complete calculation for the ${\cal N}=4$
SYM/Type IIB gravity correspondence is given in \cite{us2}). The
leading order result in the large-N expansion was first found by
\cite{henningson}, but at subleading order there are contributions
from all the Kaluza-Klein modes of supergravity, the contribution
of each supergravity field on AdS being given by a universal
formula.

In our calculation of the anomaly Dirichlet boundary conditions
were assumed for all of the bulk fields (there are no fields with
masses in the range for which Neumann conditions are admissible).
It would be interesting to consider compactifications (such as
Type IIB supergravity on $AdS_5\times T^{1,1}$) for which there
are masses allowing Neumann conditions.

In this section we will extend our result for the Weyl anomaly to
theories with double-trace perturbations (or Neumann boundary
conditions for some of the bulk fields). As explained in
\cite{us2}, the Weyl anomaly is given by the response of the free
energy to a Weyl scaling of the boundary metric, which is
equivalent to scaling the cutoff $r_0$, so that the contribution
of a bulk scalar to the anomaly is

\be \int d^dx\sqrt{\hat g}\delta{\cal A}={\partial\over\partial
r_0}F=\hf{\rm Tr}\Gamma, \ee

where we used the Schr\"odinger equation (\ref{gam}). The
expansion of $\Gamma$ in powers of the differential operator gives

\be \Gamma=\sum_{n=0}^\infty b_n(r_0) \left(\Box+{(d-2){\hat
R}\over4(d-1) }\right)^n\,,\label{ex} \ee with \be
b_0=-\sqrt{m^2+{d^2\over4}} \ee and $b_n\to0$ as $r_0\to\infty$
for all $n\ne0$.

The functional trace is regulated with a Seeley-de Witt expansion
of the heat-kernel

\be {\rm Tr}\, \Gamma=\sum_{n=0}^\infty b_n(r_0)\,\left(-{\pd
\over \pd s} \right)^n {\rm Tr}\,\exp
\left(-s\left(\Box+{(d-2){\hat R}\over4(d-1) }\right)\right), \ee
\be {\rm Tr}\,\exp \left(-s\left(\Box+{(d-2){\hat R}\over4(d-1)
}\right)\right) =\int d^d x\,{\sqrt\hg}{1\over (4\pi s)^{d/2}
}\left(a_0+s\,a_1(x)
+s^2\,a_2(x)+s^3\,a_3(x)+..\right),\label{sdw} \ee

with $s$ small. Assuming an even-dimensional boundary, in the
limit $s\to0$ and $r_0\to-\infty$ the only surviving contributions
are from $a_0,a_1,\ldots a_{d/2}$. The coefficients of all but
$a_{d/2}$ diverge, and can be cancelled by adding counterterms to
$F$, but the finite contribution proportional to $a_{d/2}$
determines the anomaly. Since $\sqrt{m^2+d^2/4}=\Delta-d/2$, we
find that

\be {\cal A}=-{\Delta-d/2\over2(4\pi)^{d/2}}a_{d/2}.\ee

All this assumed Dirichlet conditions for the canonical bulk
field, but we would like to know if Neumann conditions give a
different result for the anomaly. To change from Dirichlet
boundary conditions to the more general condition
$\dot\p=\lambda\p$, we add to (\ref{wf}) the boundary term
$\exp(-{\alpha\over2}\int\hp^2)$ and integrate over $\hp$, giving
the determinant $\det^{-1/2}(\Gamma-\lambda)$. Since $\phi$ has
the asymptotic form
$\alpha\tau^{\Delta-d/2}+\beta\tau^{d/2-\Delta}$, what we called
Neumann boundary conditions (diagonalising $\beta$) correspond to
$\lambda=d/2-\Delta$.

We have

\be \det(\Gamma-\lambda)=e^{-2\delta F}, \ee where $\delta F$ is
the correction to the free energy. Thus using (\ref{gam})

\be {\partial\over\partial r_0}\delta F={\rm
Tr}\left(-{\partial_{r_0}\Gamma\over2(\Gamma-\lambda)}\right)={\rm
Tr}\left(-{\Gamma^2-z^{2}e^{-\rho}\left(\Box+{(d-2){\hat
R}\over4(d-1)
}\right)-M_r^2\over2(\Gamma-\lambda)}\right).\label{acr}\ee

As $r_0\to\infty$ $\Gamma\to-M_r-{1\over 2
+2M_r}\left(\Box+{(d-2){\hat R}\over4(d-1) }\right)z^{2}+\ldots$
and for $\lambda\ne -M_r=d/2-\Delta$ (\ref{acr}) tends to zero.
For the specific value $\lambda=d/2-\Delta$, however, it tends to
Tr$(-1)$, giving a correction to the anomaly,

\be \delta{\cal A}=-{1\over(4\pi)^{d/2}}a_{d/2}.\label{res}\ee

Notice that for generic mixed boundary conditions there is no
correction to the anomaly.

From the point of view of a double-trace perturbation, as we go
from the UV (Dirichlet conditions and $f=0$) to the IR (Neumann
conditions and $f=\infty$) (\ref{res}) implies that the central
charge (as defined in \cite{cardy}) is decreased by 1. This is in
accordance with the c-theorem, which predicts that $c_{UV}>c_{IR}$
\cite{cardy, warner}. The correction to the anomaly applies even
in the case $\Delta=d/2$. In other words there is a non-zero flow
for bulk fields with masses saturating the Breitenlohner-Freedman
bound, in contradiction with the assumption to the contrary made
in \cite{gubser}. This explains the discrepancy with that result.
From the gauge theory side the result of \cite{gubser} was
reproduced in \cite{kg}, but the zeta-function regularisation used
there is tantamount to making the same assumption, since it is
based on an expansion in odd powers of $\Delta-d/2$.

The result (\ref{res}) can be extended to all the fields of
Supergravity by using the results of \cite{us2} in which the
Schr\"odinger equations for bosonic and fermionic fields of higher
spin were reduced to the same form as the scalar field equation.
This will be discussed in a forthcoming publication.

\end{document}